# Compilation Forking: A Fast and Flexible Way of Generating Data for Compiler-Internal Machine Learning Tasks

Raphael Mosaner[a], David Leopoldseder[b], Wolfgang Kisling[a], Lukas Stadler[c], and Hanspeter Mössenböck[a]

a   Johannes Kepler University, Linz, Austria
b   Oracle Labs, Vienna, Austria
c   Oracle Labs, Linz, Austria

**Abstract**   Compiler optimization decisions are often based on hand-crafted heuristics centered around a few established benchmark suites. Alternatively, they can be learned from feature and performance data produced during compilation.

However, data-driven compiler optimizations based on machine learning models require large sets of quality data for training in order to match or even outperform existing human-crafted heuristics. In static compilation setups, related work has addressed this problem with iterative compilation. However, a dynamic compiler may produce different data depending on dynamically-chosen compilation strategies, which aggravates the generation of comparable data.

We propose *compilation forking,* a technique for generating consistent feature and performance data from arbitrary, dynamically-compiled programs. Different versions of program parts with the same profiling and compilation history are executed within single program runs to minimize noise stemming from dynamic compilation and the runtime environment.

Our approach facilitates large-scale performance evaluations of compiler optimization decisions. Additionally, compilation forking supports creating domain-specific compilation strategies based on machine learning by providing the data for model training.

We implemented compilation forking in the GraalVM compiler in a programming-language-agnostic way. To assess the quality of the generated data, we trained several machine learning models to replace compiler heuristics for loop-related optimizations. The trained models perform equally well to the highly-tuned compiler heuristics when comparing the geometric means of benchmark suite performances. Larger impacts on few single benchmarks range from speedups of 20% to slowdowns of 17%.

The presented approach can be implemented in any dynamic compiler. We believe that it can help to analyze compilation decisions and leverage the use of machine learning into dynamic compilation.

## The Art, Science, and Engineering of Programming

**Compilation Forking**

# 1 Introduction

Dynamic optimizing compilers are complex software systems, requiring broad domain expertise to grasp the impacts of single optimization transformations on the overall program performance. Compiler experts typically fine-tune such optimization decisions based on (micro-)benchmarks, where the compilation process and the runtime performance are reproducible and stable. This results in a large set of compiler heuristics which guide the compilation process for real-world programs, by selecting the optimization parameter values to be used. For example, Leopoldseder et al. [29] introduce heuristics for estimating code size and performance impacts of compiler optimizations which are used to choose loop unrolling factors. These heuristics are typically one-size-fits-all and are optimized for code patterns found in the benchmarks used for evaluation. Therefore, different users compile their programs with the same heuristics, which might not be optimal for the particular domain or program. Manually creating heuristics for different domains or programs is often infeasible.

Data-driven approaches—often using machine learning—have been shown to outperform human-crafted heuristics for compiler optimizations [47, 2, 27]. However, the problem of generating appropriate data for training sophisticated models is often hard to solve [13]. First, compiler flags are typically not per compilation but global, which requires creating minimal programs to capture the impact of a compiler flag in isolation. Second, in a dynamic runtime, compilations are not deterministic and subject to profiling data or memory usage. Thus, it is often infeasible to create a setup for a dynamic compiler, where the impact of a single compilation parameter can be measured. Such measurements however, are required to train a machine learning model. Previous research has come up with ways to find optimal compilation plans for single functions in terms of peak performance [5, 27, 24]. However, these approaches are neither generally applicable nor suited to be used in large-scale data generation for dynamic compilers. For arbitrary methods, there are two noise factors which hamper the inference of compiler knowledge based on performance analysis: Firstly, *compilation noise*, where the same method can be compiled in different ways. This is often the case in a dynamic compiler, where compiler threads run in parallel to the executed program and profiling information is used by the compilation process. Secondly, *usage noise*, where—due to different usage scenarios, parameters or global values—a method's execution time might have high variance. While related work ignores one or both of these noise factors, we take both into account to get more reliable measurements.

We propose *compilation forking*, a technique for extracting optimization performance data in a dynamic compilation system for arbitrary programs. Its core idea is to fork method compilations at points of interest to ensure a common profiling and compilation history. A point of interest is a point in an ongoing compilation, for example right before loop unrolling is applied and the compiler has to choose one of multiple unroll factors. Starting from this common past, each compilation is completed with a different optimization parameter, and all resulting method versions are executed alternatingly. This eradicates compilation noise, up to the point of interest, and averages out usage noise in the long run. The generated data can be used to facilitate quality analysis of compiler optimizations, which can be seen in Figure 1. Therein, the hand-crafted loop peeling heuristic in the GraalVM [51] compiler is analyzed using compilation forking. The x-axis shows bins for the relative impact on execution time when applying the loop peeling transformation. It is scaled logarithmically. Therefore, values smaller than one indicate a slowdown caused by peeling and values larger than one a speedup. Outliers are cut off in both directions. The y-axis shows the number of loop peeling transformations in the respective bins e.g., the bin at the x-value 10 holds the number of peeling-transformations that led to a speedup of factor 10. The different grey-scales connect the measured performance impacts with the peeling decision that GraalVM would have made: True positive and true negative counts indicate that the GraalVM compiler correctly applied or neglected a loop peeling transformation. False negatives indicate that the compiler did not apply loop peeling although it would have produced a speedup. False positives indicate that the compiler applied loop peeling although it resulted in a slowdown. Large impacts of peeling result from interference with other optimizations (like vectorization) and rare patterns, where peeling enables removing whole loops. Compilation forking allows for such an assessment of compilation decisions for arbitrary programs.

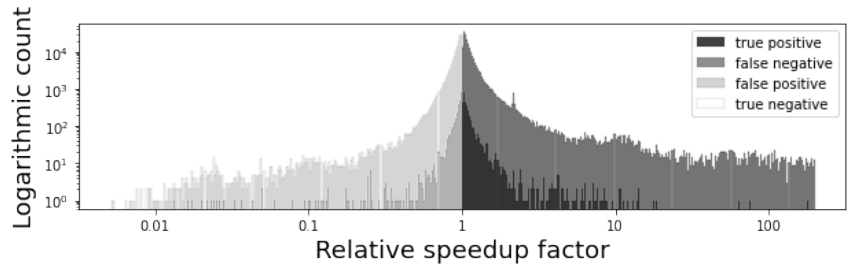

**Figure 1** Loop peeling in the GraalVM compiler.

Additionally, compilation forking can be used to train machine learning models in order to replace human-crafted heuristics, which we show in Section 5. In summary, this paper contributes the following:

- Compilation Forking: a novel approach for comparing local optimization decisions in a dynamic compiler under the same conditions on arbitrary programs.
- An elaborate performance measurement strategy that takes compilation noise, usage noise, and OS noise into account.
- A case study where compilation forking data is used to train machine learning models to predict loop optimization parameters.
- An evaluation in which a dynamic highly-optimizing production compiler is matched by learned models for loop optimizations.

The remainder of this paper is structured as follows. Section 2 gives an overview on machine learning in compilers and related work on data generation. Section 3 outlines the general process of compilation forking with Section 4 going into details on implementation specifics. Section 5 summarizes case studies where we trained machine learning models using data which is generated by compilation forking. Finally, in Section 6, we evaluate compilation forking in terms of performance and code size impact. Additionally, we evaluate our machine learning models, to show that compilation forking indeed produces high-quality data.

## 2 Background

In this section, we briefly provide an overview on machine learning in compilers in general. Subsequently, we point towards related work on data generation for machine learning problems in compilers.

### 2.1 Machine Learning in Compilers

Over the past decades, machine learning models have been shown to outperform hand-crafted compiler heuristics [19, 47, 2, 27]. Predominantly, learned models aim to improve the peak performance of compiled programs by guiding the compilation process. Other success metrics such as memory usage or code size have been optimized in the past as well [10, 11]. However, decreased memory pressure in modern hardware has led to them being neglected in more recent literature [2], apart from few exceptions [34].

Machine learning in compilers originates from iterative compilation [5, 27]. Its idea is to repeatedly compile programs with different sets of compiler parameters or a different order of compiler phases. There is extensive work on finding the best global compiler flag setup or phase plan for given programs by using iterative compilation [47, 2]. Furthermore, it can be used for creating a gold standard for evaluating other models or heuristics [19]. Auto-tuning frameworks, like OpenTuner [1], focus on reducing the state space for iterative compilations to converge more swiftly on a near-optimal program compilation. For establishing a more general relationship between source programs and beneficial compilation parameters, source code has been abstracted to descriptive features [2, 47, 27]. These features were then used in machine learning models, ranging from decision trees [33, 43], to genetic algorithms [44, 10, 7, 45], support vector machines [44, 36, 41] and neural networks [43, 32, 12, 6, 24]. Developments in the area of deep neural networks have reduced the need for extensive feature engineering and pre-processing by having these tasks taken over by the model itself [12, 27]. Thus, the traditional offline learning pipelines have adopted neural networks as their main instrument. Recently, research towards online learning in compilers has increased. Therein, trained models are improved at run time by rewarding advantageous decisions [23, 21].

Many different compiler optimizations have been investigated with machine learning models. There are models for performing inlining [43, 7] or vectorization decisions [21], for finding loop unrolling factors [33, 44] or for addressing the phase ordering [23] or skipping [24] problems. More general models aim towards predicting the performance impact of arbitrary compiler optimizations, rather than directly predicting the most beneficial optimization decision [14, 32].

In contrast to the reportedly good results in research, to the best of our knowledge, none of the optimizing compilers in HotSpot, JavaScript V8 [46] or GraalVM [51] are using machine learning to make decisions during dynamic compilation. We believe that compiler experts back off from employing machine learning black boxes in compilers due to the concern of degrading understandability and maintainability. Thus, despite successful implementations in research compilers such as Jikes RVM [7] or MILEPOST GCC [19], machine learning is not found in dynamic production compilers.

**2.2 Related Work**

In principle, our work is related to iterative compilation [5, 27] and multi-versioning [9, 52, 26]. However, the goal of compilation forking is not to produce near-optimal performance of methods by repeated compilation, but to infer impacts of local optimization decisions for later analysis. To the best of our knowledge, past work mainly optimized global compiler flags in iterative compilation for whole programs [19, 44, 43]. For instance, Stephenson et al. [44] re-run benchmarks multiple times with the loop unrolling factors set globally to a particular value. They instrument each loop and compare the performance of loops from different compilations to each other. We investigate compilation parameters locally and execute different versions alternatingly in the same program run for better stability.

Fursin et al. [18] created an approach for supporting iterative compilation by compiling multiple versions of each function, with differing compilation parameters. However, their work focuses more on finding performance stability patterns, which allows for making iterative compilation more feasible by reducing the evaluation time of different versions. They used the EKOPath compiler, which can be used for statically compiling C, C++ or Fortran programs. Our approach uses a dynamic compilation system with deoptimization [49] which enables an even more transparent usage by switching back to one favored version after data generation without interrupting the program.

Multi-versioning [9, 52, 26] is an approach related to iterative compilation, where multiple versions of a function or code snippet are deployed into an executable. At run time, the code which is best optimized towards the current input is selected. Our goal is neither to have multiple versions deployed into a production system, nor to create an optimal version during forking. Rather, we create different (non-optimal) versions to investigate the impact of local optimizations on the function performance in a fine-grained and consistent manner. Based on the gathered information, either improvements in the human-crafted heuristics can be deployed or machine learning models replacing the human-crafted heuristics.

Another research related to our work has been conducted by Sanchez et al. [41] in the IBM Testarossa JIT compiler. Starting from the conventional compilation plan, they successively remove optimization phases to compile different versions of a function. After a number of invocations of the function, re-compilation is triggered and another version is compiled with a different compilation plan. By measuring execution times based on processor timestamp counters they try to learn the best compilation plan for a given method. While their approach was successful for reducing start-up time, the overall throughput was reduced for most benchmarks. We hypothesize that their use of overall method execution time (including callees) instead of self time (excluding callees), the small number of training data points, and the goal of learning a whole phase plan for an already optimized compiler were the main reasons for not outperforming the baseline compiler. Thus, we aim to inspect differences arising from compilations in isolation, with compilation forking enabling us to start from a common past.

Lau et al. [26] also use multi-versioning in the IBM V9 compiler to determine the fastest of two versions of a function with statistical significance. They discovered that a large number of function invocations is necessary to correctly identify speedups between different versions: they argue that at least 1000 invocations are necessary to reason about differences in the 10% ranges. Second, they verified that operating system and usage noise indeed average out in the long run. We could confirm both findings in our own work, yet discovered that the longer an application runs and the more observations are recorded the better the overall data quality becomes. In contrast to their work, our system is capable of creating multiple forks per function for all compiled functions of a program run. This allows our approach to automatically extract observations, without any interaction or function pre-selection by the user. This holistic approach is - in part - enabled by our more fine-grained timestamping instrumentation. We extract self time instead of total time because we want to support all features of a dynamic execution system like the JVM, this includes deoptimization, native calls, garbage collections and transitive function calls. For all of the above "function exits" their total time does not contribute to the actual time spent in a single function with respect to function local optimization decisions. This means any knowledge system built upon this data would be biased towards total time, effectively prohibiting us to later learn any correlation between function local features and the performance of a single compiled, non-exited, piece of code. The use of total time limits the scalability of the approach shown in [26]. Additionally, we do not rely on background worker threads, which might have impacts on the measurements in a meta-circular environment.

More recently, Cummins et al. [13] addressed the problem of too few available data from a different angle compared to our approach. In essence, they use data augmentation by synthesizing programs to enlarge the set of benchmark data and thus the set of data points being usable for machine learning in compilers. They reported improvements of over 25% when training a predictive model with the extended, synthesized data. While they generate artificial benchmarks for collecting data, our approach enables executing arbitrary user programs for collecting data.

In their recently published work, Mpeis et al. [35] investigated the minimal state of a program to be captured online to later replay that programs under changed conditions offline. While our goals to conduct consistent performance measurements for changed optimization parameters are similar, the approaches are fundamentally different. Additionally, our work is not restricted to side-effect and I/O-free programs.

The span of related work, ranging back more than two decades is still present in recent research and shows that (1) *automated* generation of performance data in a (2) *dynamic compilation* system for (3) *local optimizations* is yet an unsolved problem.

## 3 Compilation Forking

We propose *compilation forking*, a technique which allows for flexible data generation during conventional program execution. Figure 2 shows an abstract depiction of our system architecture. It combines the process of compilation forking with the use case of training a machine learning model from the extracted data. The key idea of compilation forking is to create multiple versions of a compilation, sharing the same compilation and profiling history, but going separate ways at a compilation decision under investigation. Figure 2 shows at the very top the compilation pipeline, where previous transformations such as *OptA* are applied identically to all forks. The center part of the figure depicts the creation of different versions, based on the number of options for *OptX* (e.g. loop peeling). In the center right, the different versions are recombined after they have passed through the remaining compilation pipeline. Each version is executed transparently until enough data is gathered for statistical analysis or machine learning (i.e. training or inference).

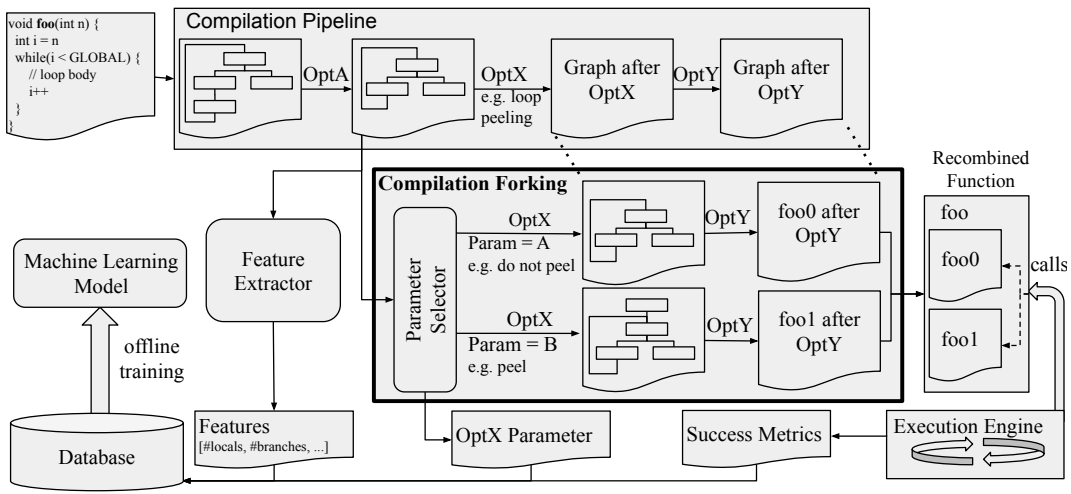

**Figure 2** System architecture. Forking happens in the *loop peeling phase*.

Executing multiple versions in a single program run has several advantages over traditional approaches which execute versions in separate runs. In the context of dynamic compilation, compiled functions depend on profiling information, timing and memory conditions. These dependencies and the whole compilation history are automatically taken into account in our approach and thus provide more comparable results compared to iterative compilation. Additionally, executing multiple versions in a single run—preferably alternating—reduces requirements regarding CPU stability and varying background tasks on the execution environment. Lastly, no additional post-processing of data from multiple runs is necessary; one program run suffices to produce consistent performance data. Compilation forking is neither restricted to being used with particular benchmarks, nor to mere execution runs for data generation.

In the following, we present a step-by-step overview of the compiler-agnostic compilation forking approach. As a running example, forking is applied in the *loop peeling* phase (c.f. Section 5.1) for function foo, shown in Figure 2. More complex steps and implementation details are explained in respective subsections of Section 4.

**Forking Point** Initially, an entry point for compilation forking has to be defined. This can be before any phase in the compilation pipeline where a compilation decision is to be made and its impact needs to be analyzed. Additionally, multiple forking points can be defined, resulting in a nested forking scheme. However, nested forking should

**Compilation Forking**

be applied with caution, i.e., only for strongly related optimizations to evaluate their interplay. Otherwise, it could lead to the initial problem of not being able to attribute performance changes to particular compilation parameters. A method is processed by all compiler phases preceding the phase at the next forking point, i.e. the forked phase, resulting in a so-called intermediate compilation.

■ **Listing 1** Fork without peeling.

```
1  void foo_0(int n) {
2      int i = n
3      while(i < GLOBAL) {
4          // loop body
5          i++
6      }
7  }
```

■ **Listing 2** Fork with peeling.

```
1  void foo_1(int n) {
2      int i = n
3      if(i < GLOBAL) {
4          // loop body
5          i++
6      }
7      while(i < GLOBAL) {
8          // loop body
9          i++
10     }
11 }
```

■ **Listing 3** Recombined forks.

```
1   void foo(int n) {
2       switch(forkControl % nrForks) {
3           case 0:
4               int i = n
5               while(i < GLOBAL) {
6                   // loop body
7                   i++
8               }
9               break
10          case 1:
11              int i = n
12              if(i < GLOBAL) {
13                  // loop body
14                  i++
15              }
16              while(i < GLOBAL) {
17                  // loop body
18                  i++
19              }
20              break
21      }
22      forkControl++
23  }
```

**Fork Creation**   At a forking point, the intermediate compilation is duplicated n times, with n being the number of compilation parameter values to be explored. In general, compilation parameter values can be either boolean, multi-class nominal or metric. Thus, the state space needs to be reduced to a manageable size. As loop peeling is a boolean decision (peel or not peel), we have to copy function foo only once to cover both scenarios for its while loop. In Section 5.3 we discuss how we used profiling information to further reduce the state space. After duplication, the forked phase is applied to each copy with the compilation parameter(s) enumerating the set of values to explore. This creates n versions of the intermediate compilation, which only differ in the parameter value for the current phase, but share the same past. For function foo, two forks—foo_0 and foo_1—are created, which are shown in Listing 1 and Listing 2. In reality, intermediate compilations would be represented in a compiler-specific intermediate representation and not in source code.

**Feature Extraction**   We use the term features as set of all *relevant* information we have about a compilation at a particular point in the pipeline. This includes both information on the compilation unit and on the optimization parameter values chosen for a fork. In a machine learning context, features are the input to a model, which produces a target value as its prediction. Feature extraction has to happen immediately before an optimization is applied.

**Compilation Pipeline**   After forking, each fork is run through the remaining compilation pipeline. Parameter values chosen in the forked phase, might have large impacts on subsequent phases. Also factors such as timing or memory usage might have an impact on the remaining phases, leading to noise in the generated data. Thus, we would interfere with the nature of a dynamic compiler if we set the remaining compilation pipeline to a fixed state.

**Success Metric Instrumentation**   After compilation of the forked function has finished, success metrics for all versions need to be extracted for evaluation. Success metrics can either be the more prevalent execution time, but also memory usage or code size. For extracting run-time metrics, we instrument the compiled function to provide the success metrics during execution. A detailed description of how we extract execution time is given in Section 4.1.

**Fork Recombination**   Eventually, all compiled forks are recombined to a self-contained function. It mimics the initially forked function by transparently dispatching to one of its versions during execution. For the running example, this is schematically depicted in Listing 3 on source code level, omitting any instrumentation for success metric or feature extraction. There are several advantages of recombining forks into a single function rather than having multiple functions at hand. Overall code size and call overhead are reduced and no undue patches to the deoptimizer and method call logic have to be made. The code size for forked functions might increase a lot, but as the control flows of different forks never merge again, no additional pressure is put on register allocation, which can use the same registers for all forks. A detailed depiction of the fork recombination process for compiler graphs is presented in Section 4.2.

**Fork Execution**   The execution of forks in the recombined function can follow any kind of order, depending on the instrumentation parameters. In our reference implementation, the forks are executed alternately, as it is indicated in Listing 3. In contrast to our approach, Sanchez et al. [41] compile a new version of a function after enough (sequential) measurements have been taken. By referring to the work of Lau et al. [26], we believe that our approach will better average out CPU frequency jitter and impacts of varying parameters. Garbage collection (GC) caused by one fork execution might impact subsequent fork executions. However, we decided to exclude GC time at safepoints in our timestamp instrumentation, as GC is hard to be attributed to certain functions. After execution the information triplet of features, success metric and compilation decision can be used for training a machine learning model or to find the *best* optimization decision ad-hoc. Using deoptimization, this *best* version can then replace the instrumented, multi-version function initially created by our forking approach.

**Requirements & Limitations**   The concept of compilation forking can be implemented in any compiler where optimizations are applied in a deterministic order. However, it is advisable that inlining and other optimizations which work across function boundaries precede the first forking point. Otherwise, the extracted success metric might be polluted by inlinees. For example, using method self time as presented in Section 4.1 would only work for optimizations after inlining. The types of possible optimizations comprise all those which can be decided based on a set of features observed at the point of an optimization. However, feature extraction and the forking point have to be defined manually for new optimizations.

**Compilation Forking**

Compilation forking is not a holistic approach, but rather allows for analyzing the impact of single or few optimization decisions, where an interplay is expected. For more holistic approaches we refer to recent related work tackling phase ordering [23] or skipping [24].

## 4 Implementation

We implemented compilation forking in the GraalVM [51] compiler, a highly optimizing dynamic compiler which is used in production on millions of devices. Our implementation is independent from source code as it directly uses Graal's graph-based intermediate program representation (IR) [17, 15]. The Graal IR consists of two directed graphs, one for control flow and the other for data flow. Nodes in the graph can either have a fixed position in the control flow (fixed nodes) or can be executed anywhere as long as data dependencies are met (floating nodes). By directly instrumenting Graal IR, we can profit from the polyglot features provided by GraalVM and its language implementation framework Truffle [50]. This enables extracting performance and feature data for any Truffle language, which facilitates training of language-specific machine learning models without additional effort. We now present more detailed implementation insights.

### 4.1 Timestamp Extraction

For most compiler optimizations, the impact on the execution time of the compiled function is of highest importance. In this section, we show our instrumentation for extracting *self time* of methods or code snippets. Self time is the time spent executing a method excluding time spent "outside" in calls or at safepoints[1] [28]. The basic idea behind this instrumentation is shown as pseudo code in Listing 4 and Listing 5. Therein, the self time of the current execution is first aggregated locally by excluding calls (or safepoints). At the end, the current execution time is aggregated to the global, thread safe storage.

■ **Listing 4**  Base function.

```
1  int bar(int n) {
2
3      int m = n + 1
4
5
6      int res = foo(n, m)
7
8      res = res * 2
9
10
11
12      return res
13  }
```

■ **Listing 5**  Pseudo instrumentation.

```
1  int bar(int n) {
2      long t = 0, ts1 = getTime()
3      int m = n + 1
4      long ts2 = getTime()
5      t += ts2 - ts1
6      int res = foo(n, m)
7      long ts3 = getTime()
8      res = res * 2
9      long ts4 = getTime()
10     t += ts4 - ts3
11     aggregate("bar", t)
12     return res
13  }
```

---

[1] Points in the program execution where the GC can be safely run. See https://openjdk.java.net/groups/hotspot/docs/HotSpotGlossary.html

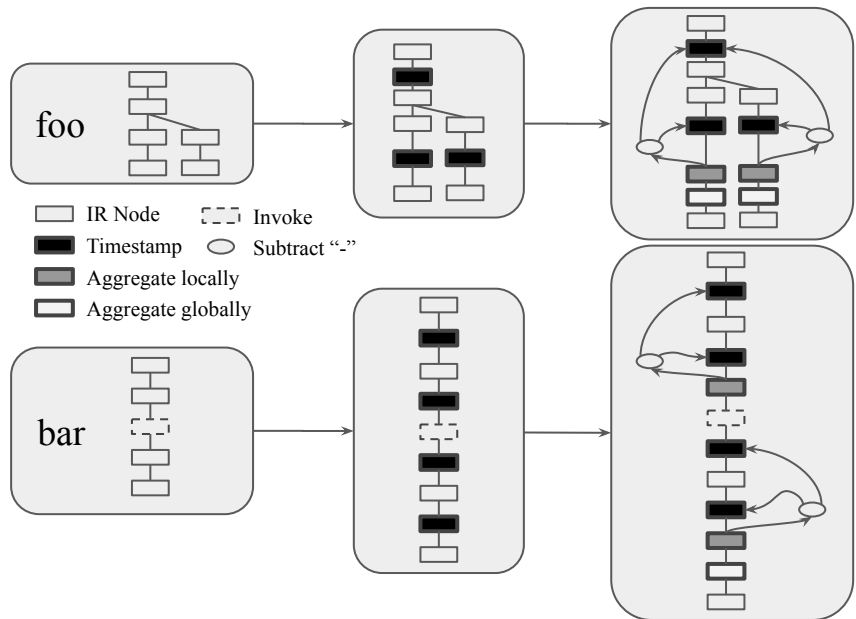

**Figure 3** Timestamp instrumentation.

The actual instrumentation is added in the graph-based compiler IR, shown in Figure 3. Rectangular nodes are fixed nodes, in the sense that their order is preserved in a scheduling process. Circular nodes are nodes for subtraction, which can be executed out of order if data dependencies are met. Timestamps (black nodes) are added after each method entry (=first) node and before each method exit (=last) node, which is shown in the first instrumentation step in Figure 3 (foo). The difference between these timestamps would correspond to the *total* execution time of the method. To extract *self* time, we add additional timestamps before and after invocations of other methods and safepoints, which can be seen in Figure 3 (bar). While the instrumentation for measuring self time is more complex, we want to stress the importance of using it instead of total time. Otherwise, optimizations in callees would impact the *measured* performance of the caller. This could lead to an optimization being falsely identified as beneficial or harmful for the caller features. In case of across-function optimizations, like inlining, we can switch to total time to capture the time spent for a whole call.

Each time snippet is calculated by subtracting the start timestamp from the corresponding end timestamps (circular nodes). These measurements are summed up locally (dark grey nodes) for calculating the current method self time. Eventually, before each control flow sink, e.g. return or exception, the current execution time is added to the aggregated self time for this fork. This is captured in the white nodes which also handle the necessary synchronization of that operation. Storing all execution timestamps for a fork at run time would be infeasible with millions of invocations of each fork. However, by storing the aggregated run time of a fork together with its invocation count, we can calculate an average execution time for each fork, similar to [26]. We assume that—enough invocations provided—different method execution times due to different values of parameters or globals will average out as proven in [26]. The rationale behind this is the fact, that an optimization can only be considered beneficial, if the *average* execution time of the optimized code is improved.

**Compilation Forking**

**Timestamps**   The timestamp values are extracted using Intel's rdtscp[2] instruction. This instruction ensures that preceding instructions are executed before acquiring a timestamp. Additionally, we emit lfence instructions after start timestamps and before end timestamps as shown in Figure 4. lfence instructions ensure that instructions scheduled after them will not be executed out-of-order before the lfence instruction is completed. Using the setup as shown in Figure 4 excludes the time for executing the instrumentation logic (e.g. when updating the aggregated local time). Thus, only the end timestamp call "pollutes" our measurement, which is the minimum unavoidable noise when timestamping. Nevertheless, the usage of lfence instructions will impact small or empty measurement regions a lot. In addition, if the performance of different forks is vastly affected by limiting out-of-order executions via lfence instructions, measurement noise might occur. Section 6.1 shows the overhead of the timestamp instrumentation for different types of functions. We use a data filtering to omit these data points from the result set.

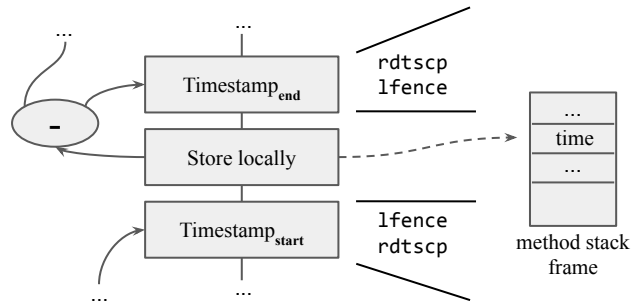

**Figure 4** RDTSCP and memory fences.

**Local Storage**   For the timestamp instrumentation a local slot in the method's stack frame is allocated via instrumentation to track self time across calls and safepoints. This ensures a thread-safe and recursion-supporting self time measurement. A measurement section can either end because of a control flow sink, e.g. return or exception, or an excluded operation such as a method call. Then, the method's stack slot value is updated by adding the difference of the timestamps from the current region.

**Global Storage**   The global storage holds the aggregated time of all method executions, which are again subdivided into all forks. It resides in the compiler and is updated once before a control flow sink is encountered. The update is managed via GraalVM's AtomicReadAndAdd operation, which maps to an xadd on Intel and ensures synchronization.

**Outlier Handling**   During feature extraction, we extract both static information and profiling information at the time of compilation. Other dynamic information such as values of parameters or globals are not extracted in our approach. However, we argue that performance differences caused by different values of parameters or globals will cancel out in the long run [26]. However, noise from operating system (OS) interference still might occur and is hard to avoid in a real-world environment. The potential sources for OS noise include scheduling, context switches, memory usage and caching. We empirically checked that this noise is not introduced by our approach or instrumentation, by experimenting with native C programs, which exhibited similar OS noise. This experiment confirmed that while most execution times are stable, outliers from OS noise can exceed the average execution time by orders of magnitude. This can especially distort results for short-running methods. To counter OS outliers, we implemented an on-the-fly outlier removal as a transparent, yet aggressive addition to our timestamp instrumentation. It compares each locally aggregated execution time to the global average execution time for the fork. Depending on an outlier threshold local times may be omitted from being added to the global counter. Consistently, the invocation counter is not increased if an outlier is detected. With this measure, outliers accounting for sometimes 10% of the total execution time could be filtered out. However, we have to point out that it is impossible to distinguish between outliers caused by OS noise and outliers from extreme usage patterns. An single invocation of a fork with a exceptionally large parameter value would therefore likely be classified as an outlier. We are also experimenting with the use of a real-time OS to investigate the outlier behavior in a fully controlled environment, which might be put to use in the future.

---

[2] https://www.intel.com/content/dam/www/public/us/en/documents/white-papers/ia-32-ia-64-benchmark-code-execution-paper.pdf

### 4.2 Fork Recombination

Fork recombination happens at the very end of the compilation pipeline. All compiler graphs originating from one or multiple forking events are merged into a combined graph resembling a pseudo-function. Figure 5 depicts the recombination process for a function where forking has been applied in the loop peeling phase. Initially, each fork is compiled on its own, including timestamp instrumentation or any other success metric measurement. Eventually, the forks are recombined by copying their graphs into different branches of a switch. This switch—shown in source code for simplicity—controls the fork execution. In Figure 5 an alternating execution of the two forks is enforced. The *forkControl* variable is stored in the compiler, as discussed in Section 4.3. During code generation, it has to be ensured that all switch-cases are aligned identically. Otherwise, varying alignment can cause reproducible performance differences.

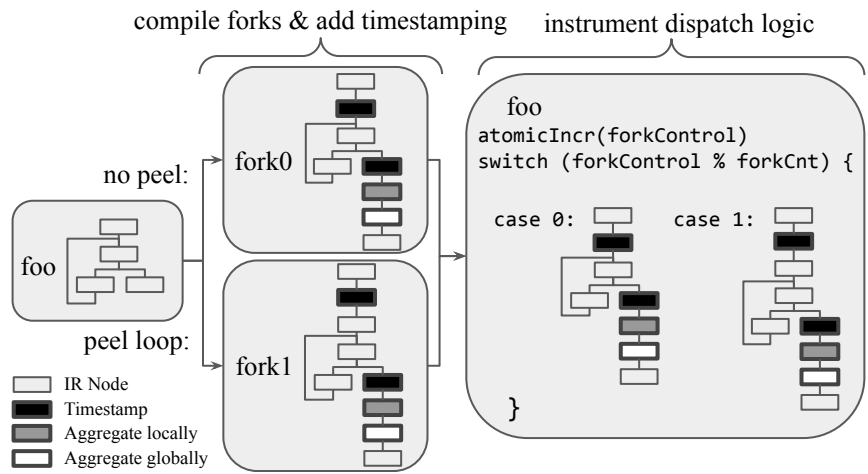

**Figure 5** Fork recombination.

## 4.3 Data Structures

To avoid pressure on the garbage collector, which could impact the program execution, we store dynamically extracted performance information in a pre-allocated array of type long. This array is static, resides in the compiler and is persisted at the time of success metric collection or at program termination, at the latest. The storage format is shown in Figure 6. *C1* denotes the first forked compilation unit, with *C1F1* to *C1Fn*

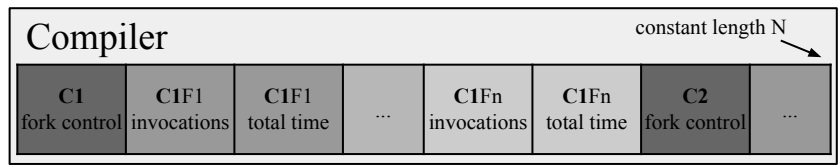

■ **Figure 6** Storage format for performance data.

summarizing all its forks. The first field for each compilation unit is the *fork control*, which is used to choose the next fork to be executed (see Section 4.2). Right after the *fork control*, we store for each fork the number of invocations and the total execution time. All array indices are compiled as constants in the instrumentation code.

## 5 Case Studies: Loop Optimizations

In this section, we show case studies on how compilation forking can be used as a flexible data generation framework for machine learning in compiler optimizations.

### 5.1 Optimizations

**Loop Peeling**   Loop peeling [3] is a transformation which moves a certain number of loop iterations in front of or behind the loop by copying the loop body. An example can be seen in Listing 1 and Listing 2. Peeling can eliminate null-checks within a loop, which are then performed only once outside the loop. In the GraalVM compiler, only the first iteration may be peeled. While the impact of loop peeling is generally considered low, our research with compilation forking showed that interference with other loop optimizations, especially vectorization, can have significant impacts which can be seen in Figure 1.

■ **Table 1** Loop feature overview.

| Category | Features |
|---|---|
| General | size; depth; isNested; #children; #backedges; #exits; isVectorizable |
| Execution | frequency; has[Exact/Max]TripCount; canOverflow |
| Nodes | #fixedNodes; #floatingNodes; #[IRNodeType] |
| Edges | #[EdgeType]IntoLoop; #[EdgeType]InLoop; #[EdgeType]OutOfLoop |
| Operands | #[object/int/float]Stamps; #[volatile/static]FieldAccesses |
| Parent | hasParent; parentSize |
| Graph | size; #loops; maxLoopDepth; #branches; #[IRNodeType] |

**Loop Unrolling** Partial loop unrolling [3] is an optimization where the loop body is duplicated a certain number of times within the loop. Accordingly, the loop stride and the loop condition are adapted to fit the enlarged loop body and thus a smaller number of iterations. Loop unrolling can reduce the overhead of loop condition checks and can produce larger basic blocks, enabling more optimizations. The number of duplications is called *unroll factor*, the choice of which is the key of this optimization.

## 5.2 Features

In our case study, we focus on two loop-related optimizations — peeling and partial unrolling — introduced in Section 5.1. These optimizations use features that describe the loop to be optimized. We decided to base our features on the Graal intermediate program representation (IR) rather than on source code. Therefore, we can profit from the GraalVM's language implementation framework Truffle [50] which enables executing Java, JavaScript, Python or LLVM [40] programs. Also other approaches [6, 36] have used IRs—mostly LLVM IR—to support multiple source languages. Table 1 gives an overview of the features we extracted for loop-related optimizations. Other optimizations, for example duplication [30], would need a different set of features to describe the program parts to be optimized. The total number of potential features is approximately 1000 before applying any filters. This large number results from the fact that we use the node counts of the different IR node types (of which there are more than 450) in the loop and the enclosing graph as features.

## 5.3 Data Generation

When using compilation forking for loop-related optimizations, we work with the assumption that loops have no impact on each other when being optimized. This means that the speedup/slowdown resulting from transforming a loop l1 is independent of a preceding or succeeding transformation of loop l2. While this is a bold assumption and might not hold in some cases, interference should be low for non-nested loops. Our assumption results in a linear dependency between state space size and number of loops per function. Currently, we reduce the number of forks by selecting the most frequently executed loops as targets for creating forks. This information is provided as part of the profiling information provided by GraalVM [16]. Altogether, one fork for

**Compilation Forking**

each loop and for each optimization parameter value is created, along with a baseline fork without any optimized loops. For loop peeling, in each fork exactly one loop is peeled. For loop unrolling, each fork has exactly one loop unrolled with one specific unroll factor of 2, 4, 8, 16 or 32. The differences between the average execution times of fork and baseline can be considered as the success metric for the optimization. We used state-of-the-art industry and research benchmark suites such as *DaCapo* [4], *DaCapo Scala* [42], *Renaissance* [39] and *Octane*[3] for generating data, as well as a micro-benchmark suite including more recent JVM features such as lambdas and streams. To minimize noise, we disabled CPU frequency scaling, hyperthreading and vendor-specific features which impact performance.

### 5.4 Data Preprocessing

The implemented outlier removal handles large outliers. Nevertheless, small outliers still result in noise in the performance measurements. Thus, we apply filters to increase the consistency of the training dataset and consequently the trained model. Simple filters reduce noise susceptibility by removing compilations where forks either do not exceed a minimum number of invocations or a minimum average execution time. We experimented with filtering out data with very small speedups or slowdowns which may likely be overshadowed by noise. Additionally, small speedups or slowdowns indicate that the optimization decision is of not much importance, as the (performance) outcome is similar in any case. By removing such data, the trained model will be forced to focus on learning the more important decisions, where performance impacts are of higher significance. Apart from success metric filters, we also applied filters to reduce the feature space, i.e., the number of inputs to the model. There are many features with little information. A feature is deemed more informative the more different feature values are found throughout the dataset. Many IR nodes appear very rarely or not at all in whole benchmark suites leading to many IR node count features being zero for most data points. We conduct a sparsity check to remove such features, which allows us to heavily reduce the feature space. For loop peeling we reduced the number of features for model training to 299 and for unrolling to 257. All features are standardized before use by subtracting their mean and dividing by their standard deviation. The number of raw and filtered data points for each benchmark suite can be found in Table 2.

### 5.5 Model Training

We evaluated a multitude of models with different filtering and hyperparameter setups by using state-of-the-art machine learning frameworks for Python: PyTorch [37] for neural networks, scikit-learn [38] for dimensionality reduction, random forests (RF) and support vector machines (SVM) as well as the XGBoost [8] framework for gradient boosting. However, shallow models such as logistic regression, decision trees, SVMs and RFs were insufficient in terms of prediction accuracy. This led us to different kinds of fully-connected neural networks (FCNNs). For the network layout, we tried small dense networks with up to 10 layers and a few million parameters to deeper residual networks with up to tens of millions of parameters. The latter have been shown to produce good results for tabular data [20]. Such residual networks with ten residual blocks and full pre-activation as described by Kaiming He et al. [31] produced the best results for peeling and unrolling. As optimizer we employed either plain Adam or AdamW, providing decoupled weight decay [25, 22]. We decided to train classification models in case of binary decisions such as peeling and a regressor to predict the speedups of different unroll factors. As loss function we used binary cross entropy (BCE) for classification and mean squared error (MSE) for regression. We scaled these losses by a function of the absolute logarithmic speedup and the aggregated execution time to give more importance to data points with a higher expected absolute speedup that have more impact on the overall benchmark time. In the models we use regularization layers and dropout to counter overfitting.

---

[3] https://github.com/chromium/octane

**Table 2** Data overview.

| Suite | peeling | | | unrolling | | |
|---|---|---|---|---|---|---|
| | raw | filtered | [%] | raw | filtered | [%] |
| DaCapo | 28928 | 23697 | 81,9 | 2901 | 2083 | 71,8 |
| DaCapo Scala | 47138 | 40455 | 85,8 | 4832 | 3770 | 78,0 |
| Renaissance | 128541 | 109945 | 85,5 | 11009 | 8154 | 74,1 |
| Micros | 243434 | 207200 | 85,1 | 30359 | 22598 | 74,4 |
| Octane | 14079 | 12949 | 92,0 | 707 | 508 | 71,9 |

We used a cross-validation approach, where we randomly divided benchmarks into groups of five. A model is trained on all data except the five benchmarks from the corresponding evaluation group. Following this approach, we trained 28 models with identical hyperparameters for peeling and unrolling each. Therefore, each model is evaluated on truly unseen data in Section 6.2. The networks were trained for 2000 to 3000 epochs with a declining learning rate every 400 epochs. While training, we used a train-validate-split (90% train, 10% validate) to get insight into the training progress. The two resulting models are summarized in Figure 7.

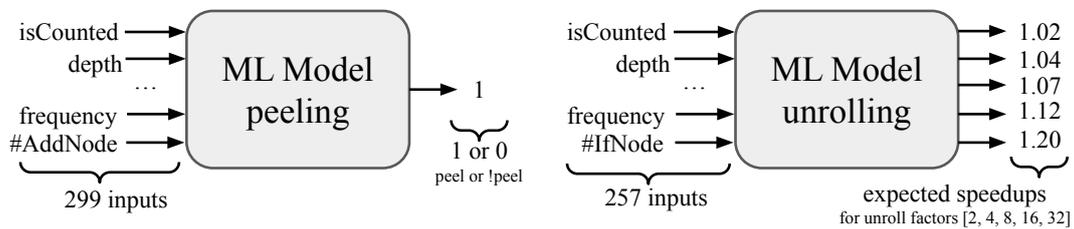

**Figure 7** ML models for peeling and unrolling.

We also experimented with overfitting on single benchmarks, where we used gradient boosting due to its faster training. For this, we used XGBoost and allowed for an ample number of estimators (500), which led to an overfitting of up to 98 percent.

**Compilation Forking**

### 5.6 Model Evaluation

Evaluating multiple models in the compiler on all benchmarks is a time-consuming task. However, standard metrics such as accuracy, precision, recall or F1-score do not reflect the quality of the estimator in terms of total performance. For quickly selecting which models to test in the compiler, we estimated the performance impact on benchmark level by employing a custom heuristic.

Based on compilation forking data, this heuristic estimates the execution time $\widetilde{t}_m$ of a method $m$ with predicted optimization parameters: We calculate the impacts of each compilation decision $d$ in $m$ compared to the baseline. Loop related optimizations yield one decision per loop and each decision can be represented by multiple forks. For example, in forked loop unrolling each unroll factor is mapped to one fork per loop. We select for each compilation decision $d$ the fork where the parameter $p$ under investigation matches the predicted value. Then, we extract the average execution time for the predicted decision $\overline{t}_d{}^p$ and calculate the difference to the baseline execution time $\overline{t}_b$. This difference denotes the expected speedup or slowdown for an optimization decision. The total execution time impact results from summing up the per-decision impacts. This total execution time impact is added to the baseline average execution time $\overline{t}_b$ and scaled by the total invocations $i$ of the method, resulting in the expected absolute execution time $\widetilde{t}_m$. This is summarized in Equation (1).

$$\widetilde{t}_m = i \left( \left[ \sum_d \overline{t}_d{}^p - \overline{t}_b \right] + \overline{t}_b \right) \quad \begin{aligned} \widetilde{t}_m &: \text{estimated average execution time of m} \\ i &: \text{total invocations of method m} \\ d &: \text{optimization decision} \\ \overline{t}_d{}^p &: \text{Average execution time} \\ &\phantom{:}\text{ with parameter p in decision d} \\ \overline{t}_b &: \text{Baseline of the method} \end{aligned} \quad (1)$$

Finally, we sum up all method execution times to estimate the per benchmark execution time when using a learned model. When exporting the default compiler decisions during forking, the same heuristic can be used to calculate the expected execution time in the current compiler. Additionally, an estimate of the best possible execution time can be made by using the minimum average execution time for each decision $d$. However, the heuristic relies on the assumption that decisions in the same method do not influence each other. It should therefore be taken as an estimator only.

## 6 Evaluation

We evaluated our approach twofold. First, we show the performance and code size impacts of compilation forking itself. Second, we show its applicability, by evaluating machine learning models, which are trained using the generated data.

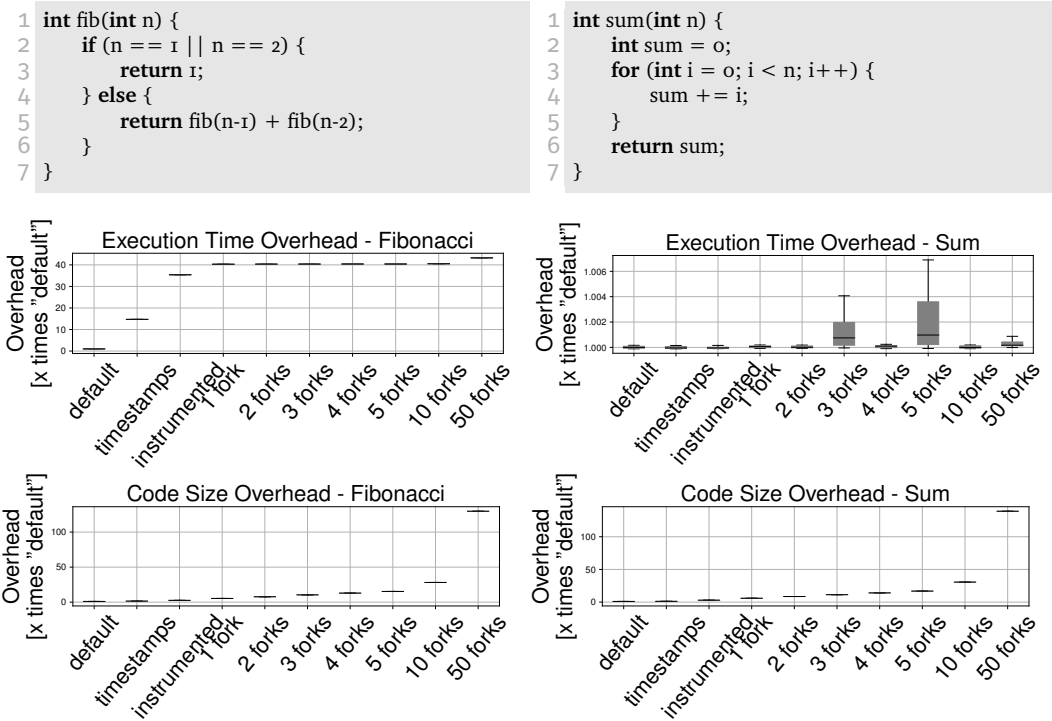

**Figure 8** Performance and code size impact of forking. Lower is better.

## 6.1 Compilation Forking

As explained previously, compilation forking allows us to compare different optimizations based on the same compilation history. The performance of optimized parts is measured by instrumentation, which is excluded from time measurements. In the presented approach, the performance data is generated and processed offline, either by compiler experts analyzing compiler performance or when training machine learning models. Therefore, we consider compilation forking to be a non-performance-critical mode, where impacts on total execution time and code size are negligible. Nevertheless, we want to show how the impact on those metrics can vary depending on the compiled code and the number of forks. Note, that we now evaluate the *overall* performance *including* the instrumentation overhead; the *measured and extracted* performance numbers are not impacted by the instrumentation overhead.

Figure 8 shows two functions and the overhead in terms of execution time and code size. The following configurations were tested, which build on top of each other: *default*: GraalVM without any changes; *timestamping*: timestamp measurement; *instrumentation*: outlier removal and invocation counting; *forks*: a number of identical forks created from the original function. The first function recursively calculates the n-th Fibonacci number. A single invocation is executed very fast, but the recursive nature leads to many calls. The second function calculates a sum in a loop. An invocation has significantly higher workload, depending on n. Each measurement was executed 50 times with 10000 calls of the function.

**Compilation Forking**

**Performance** Figure 8 shows that most overhead is introduced by timestamping and instrumentation. The relative overhead of timestamping becomes more costly the smaller the function is and the more calls have to be excluded in the self time instrumentation. Thus, a slowdown factor of 16 is encountered for function *fib*, but no overhead for function *sum* with its long running loop. The instrumentation overhead for outlier removal and invocation counting happens only once per method and adds a constant overhead. Forking introduces a constant slowdown for the dispatch logic, which is relatively larger in smaller functions. This has also been encountered in related work [18] and is the reason why smaller functions are removed from the data set. Creating multiple forks does not impact the execution time. Only in the configuration with 50 forks a small slowdown is measured for both examples. We assume that this is because of the immensely increased code size which might impact caching.

**Code Size** The impact on code size depends on the number of inserted timestamps and the number of instrumented control flow sinks. Regarding the number of forks, it follows a linear pattern, as can be seen in Figure 8.

### 6.2 Learned Compiler Optimizations

We replaced the hand-crafted heuristics in the GraalVM compiler—currently one of the highest-optimizing Java compilers[4]—with our trained models to investigate two hypotheses: First, that compilation forking does not distort the program execution and that it yields consistent performance results. Second, to investigate how data-driven optimization approaches perform against hand-crafted heuristics with years of fine-tuning and compiler expertise.

Our evaluation compares the GraalVM compiler with a compiler version using the learned predictors. Each evaluation replaces only one loop optimization heuristic by a learned model to ensure better comparability. We measured run time, code size and compile time with the former being of most interest as it has been used as a success metric in model training.

Each optimization is evaluated on five benchmark suites, introduced in Section 5.3, executed for at least ten times per configuration. To ensure a separation of training and test data, we deployed multiple models, where each model is tested on five benchmarks not used in its training. Due to space limitations, we summarize the performance results per benchmark suite for peeling in Table 3 and for unrolling in Table 4. The tables contain the number of benchmarks in each suite, the number of benchmarks where a speedup or a slowdown has been encountered, along with a maximum speedup and slowdown. For calculating speedups and slowdowns, we aggregate benchmark runs by calculating geometric means. To ensure statistical significance, we present the number of significantly faster or slower benchmarks per suite (#sig) using a Wilcoxon signed-rank test [48] for unpaired data. We avoided a standard t-test as we cannot ensure that our data is normally distributed due to noise in the executions.

---

[4] https://renaissance.dev/

For most learned models we can see comparable or slightly worse performance than for the highly-tuned GraalVM heuristics in terms of geometric means for the whole benchmark suite run time. The learned peeling strategy summarized in Table 3 seems to peel more often, as the overall code size and compile time are increased. In single benchmarks, both peeling an unrolling achieve speedups of up to 20% but also slowdowns of up to 17% .

**Table 3** Loop peeling evaluation. Lower is better for geometric means comparison.

| Suite | | Speedup | | | Slowdown | | | Geometric Means ML vs. Heuristics | | |
|---|---|---|---|---|---|---|---|---|---|---|
| Name | # | # | #sig | max% | # | #sig | max% | RunTime | CodeSize | CompTime |
| DaCapo | 9 | 2 | 1 | 0.698 | 7 | 5 | 6.165 | 1.020 | 1.203 | 1.336 |
| DaCapo Scala | 12 | 1 | 0 | 0 | 11 | 8 | 8.023 | 1.018 | 1.138 | 1.207 |
| Renaissance | 25 | 9 | 1 | 8.381 | 16 | 8 | 10.231 | 1.003 | 1.213 | 1.295 |
| Micros | 74 | 38 | 17 | 20.104 | 36 | 16 | 15.747 | 0.997 | 1.182 | 1.222 |
| Octane | 14 | 5 | 3 | 5.668 | 9 | 5 | 2.439 | 0.998 | 1.178 | 1.283 |

**Table 4** Loop unrolling evaluation. Lower is better for geometric means comparison.

| Suite | | Speedup | | | Slowdown | | | Geometric Means ML vs. Heuristics | | |
|---|---|---|---|---|---|---|---|---|---|---|
| Name | # | # | #sig | max% | # | #sig | max% | RunTime | CodeSize | CompTime |
| DaCapo | 9 | 1 | 0 | 0 | 8 | 4 | 17.637 | 1.039 | 1.071 | 1.150 |
| DaCapo Scala | 12 | 5 | 2 | 4.467 | 7 | 3 | 5.890 | 1.003 | 1.036 | 1.078 |
| Renaissance | 25 | 9 | 1 | 18.337 | 16 | 4 | 11.598 | 1.008 | 1.023 | 1.081 |
| Micros | 74 | 33 | 3 | 9.718 | 41 | 6 | 8.463 | 1.001 | 1.061 | 1.104 |
| Octane | 14 | 4 | 1 | 0.765 | 10 | 6 | 5.940 | 1.011 | 1.071 | 1.138 |

We also analyzed the potential gain of optimizations by explicitly overfitting models on single benchmarks. Detailed results for loop peeling in the Octane benchmark suite are shown in Figure 9. It indicates that for many benchmarks, significant speedups can be achieved by optimizing loop peeling, which is often considered less important. However, benchmarks like Richards or Raytrace show slowdowns compared to the default GraalVM which indicates that the data is not clearly distinguishable using our features. This was already foreshadowed in the training process, where accuracy for Richards only converged at 73%.

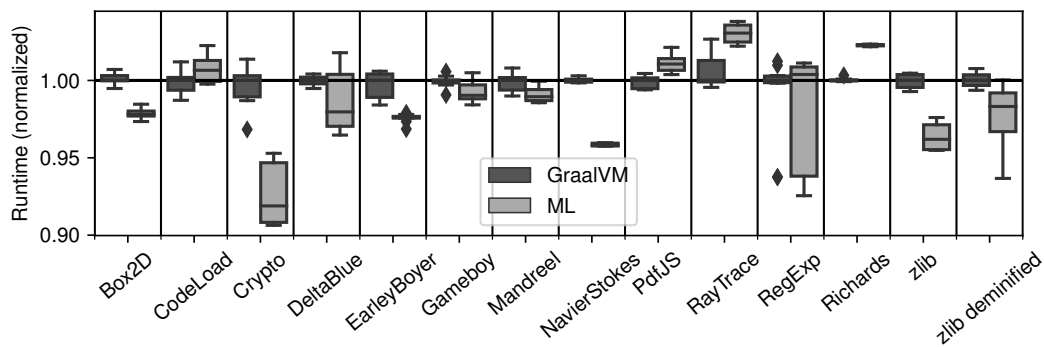

**Figure 9** Octane peeling (overfitted). Lower is better.

**Compilation Forking**

Taking into account that the GraalVM heuristics are tuned towards these benchmarks, we argue that the performance of the trained models supports our major claim: High-quality performance data can be generated using compilation forking, which can facilitate creating machine learning models that match existing compiler heuristics. Additionally, we found flaws in the GraalVM heuristics for several benchmarks and could point compiler experts to them.

## 7 Conclusion

In this paper, we presented compilation forking - a method which brings back iterative compilation into modern times for dynamic compilers. It allows for comparing different optimization decisions in dynamic compilers based on a common compilation history for arbitrary programs. Instead of re-compiling individual functions and re-starting the surrounding benchmark program, different versions of functions are compiled and executed all in one run. We handle uncertainties in the dynamic compilation pipeline, by forking a compilation at a point of interest. We execute different forks alternatingly, to create a statistically representative average over function calls with different values of parameters or globals. Thus, we claim to be able to assess the impact of single compilation decisions accurately. Compilation forking itself is compiler-agnostic. However, our implementation within the GraalVM compiler allows for a novel level of programming-language-agnostic applicability. We use its graph-based compiler-internal program representation for extracting program features. This enables generating data for any language, supported by GraalVM's Truffle language implementation framework.

To verify the quality of our generated data, we trained several machine learning models for replacing compiler heuristics for loop peeling and unrolling. Achieving similar performance to one of the fastest JVMs supports our claim that compilation forking indeed can be used to extract quality performance data and to consistently compare different optimization decisions in a dynamic compiler Speedups of up to 20% for single benchmarks unveiled flaws in GraalVM's heuristics for particular code patterns. In future work we will shift the focus on improving our machine learning models further, by employing techniques such as graph neural networks to better capture graph-based IRs. Furthermore, compilation forking itself provides opportunities for further research: We plan to investigate the interplay of optimizations by employing nested forking. Besides that, deoptimization enables us to connect data generation by compilation forking with using learned or updated ML models in single runs.

**Acknowledgements**    This research project is partially funded by Oracle Labs.

**Compilation Forking**

**About the authors**

**Raphael Mosaner** is a PhD student at the Johannes Kepler University in Linz, Austria. Contact him at raphael.mosaner@jku.at.

**David Leopoldseder** david.leopoldseder@oracle.com.

**Wolfgang Kisling** wolfgang.kisling@jku.at.

**Lukas Stadler** lukas.stadler@oracle.com.

**Hanspeter Mössenböck** is a professor at the Johannes Kepler University in Linz, Austria. Contact him at hanspeter.moessenboeck@jku.at.